\documentclass[aps,pre,twocolumn,showpacs,superscriptaddress,longbibliography]{revtex4-1}
\usepackage{graphicx}
\usepackage{amsmath}
\usepackage{amssymb}
\usepackage[pdftex,bookmarks,colorlinks,breaklinks]{hyperref}  
\hypersetup{linkcolor=red,citecolor=blue,filecolor=dullmagenta,urlcolor=blue} 
\hypersetup{
 pdfauthor={D.V. Savin, M. Richter, U. Kuhl, O. Legrand, F. Mortessagne},
 pdftitle={Fluctuations in an established transmission in the presence of a complex environment},
 pdfkeywords={}, pdfsubject={}, pdfcreator={}, pdflang={English} }

\providecommand*{\HGOE}{\ensuremath{H_\mathrm{\,bg}^\mathrm{(goe)}}}
\providecommand*{\Heff}{\ensuremath{\mathcal{H}_\mathrm{eff}}}
\providecommand*{\Gammadown}{\ensuremath{{\Gamma_{\downarrow}}}}
\providecommand*{\Gammazero}{\ensuremath{{\Gamma_{0}}}}
\providecommand*{\Gammaabs}{\ensuremath{\Gamma_\mathrm{abs}}}
\providecommand*{\Pzero}{\ensuremath{\mathbb{P}_{0}}}
\providecommand*{\Pzeroint}{\ensuremath{\mathbb{P}_{0}^{\mathrm{(int)}}}}
\providecommand*{\Pgamma}{\ensuremath{\mathcal{P}_{\gamma}}}
\providecommand*{\Pgammazero}{\ensuremath{\mathcal{P}_{0}}}

\providecommand*{\Prefletagamma}{\ensuremath{\mathcal{P}^\mathrm{(refl)}_{\gamma}}}
\providecommand*{\realizations}{\ensuremath{5{\times}10^4}}
\providecommand*{\eps}{\ensuremath{{\varepsilon_{0}}}}

\begin{document}

\title{Fluctuations in an established transmission in the presence of a complex environment}

\author{Dmitry V. Savin}
\affiliation{
Department of Mathematics, Brunel University London, Uxbridge UB8 3PH, United Kingdom}

\author{Martin Richter}
\author{Ulrich Kuhl}
\author{Olivier Legrand}
\author{Fabrice Mortessagne}
\affiliation{
Institut de Physique de Nice, Universit\'{e} C\^{o}te d'Azur, CNRS, 06100 Nice, France}

\date{\today}

\begin{abstract}
In various situations where wave transport is preeminent, like in wireless communication, a strong established transmission is present in a complex scattering environment. We develop a novel approach to describe emerging fluctuations, which combines a transmitting channel and a chaotic background in a unified effective Hamiltonian. Modeling such a background by random matrix theory, we derive exact non-perturbative results for both transmission and reflection distributions at arbitrary absorption that is typically present in real systems. Remarkably, in such a complex scattering situation, the transport is governed by only two parameters: an absorption rate and the ratio of the so-called spreading width to the natural width of the transmission line. In particular, we find that the established transmission disappears sharply when this ratio exceeds unity. The approach exemplifies the role of the chaotic background in dephasing the deterministic scattering.
\end{abstract}
\maketitle

\section{Introduction}
\label{sec:introduction}

In many applications ranging from electronic mesoscopic or quantum devices~\cite{bee97,alh00,mel04b} to telecommunication or wireless communication~\cite{tul04,cou11}, transmission and transport are the main focus of studies. Here, either the system or its excitation is designed to give a high transmission at the working energy (or frequency) $\eps$. To guarantee the functionality of such devices, fluctuations in transmission induced by a complex \emph{environment} are of crucial interest. Such variations might be introduced by uncertainties of the production process of electronic devices or by real-life changing environment like in wireless communication. In the latter case, e.g., one is interested in a stable communication that has strong transmission guaranteeing large signal to noise ratios as well as high data transfer rates~\cite{ges00,mat10}. This is often implemented via multiple-input multiple-output (MIMO) systems \cite{big07,bli13,kara16}, which control the excitations thus giving rise to a special basis, where an \emph{established} transmission is induced at $\eps$. In electronic devices like quantum dots or wells, similar established transmission is put forward in the design of the system and placement of the leads~\cite{mar12}.

The analysis of complex quantum or wave systems often relies on predictions obtained for fully chaotic dynamics ~\cite{mel04b,haa01b}. Random matrix theory (RMT) has proved to be extremely successful in describing universal wave phenomena in such systems~\cite{meh04,guhr98}. The canonical examples are spectral and wave function statistics, including their many experimental verifications~\cite{stoe07b}. When combined with the resonance scattering formalism \cite{mah69}, RMT offers a powerful approach \cite{ver85a,sok89,fyod97} to describe universal statistical fluctuations in scattering, see  \cite{mitc10,fyod11ox} for recent reviews. The approach is also flexible in incorporating real world effects, like finite absorption, providing the non-perturbative theory \cite{fyod05,kuma13} to account for statistical properties of complex impedances \cite{hem05a,hemm06}, transmission and reflection coefficients~\cite{kuhl05,kuhl05a} observed in various microwave cavity experiments~\cite{die10b,kuh13,grad14}. From the theoretical side, non-universal aspects related to a deterministic part in scattering are usually removed from the very beginning by means of a certain procedure~\cite{enge73}, yielding a new scattering matrix which becomes diagonal on average \cite{ver85a}. This, however, cannot be applied in the present case where fluctuations in a transmitting channel, characterised by the essentially non-diagonal deterministic $S$ matrix, are the main point of interest.

In this work, we propose a non-perturbative approach to quantify fluctuations in the established transmission mediated by a single energy level that is coupled to a complex environment modeled by RMT. The model in its original formulation goes back to nuclear physics \cite{Bohr}, giving rise to the well-known formalism of the strength function that has a rich history of various applications \cite{harn86,soko97,gu99,sav03b,soko10,zele16}. Nevertheless, a complete characterisation of fluctuations in the transmission for such a model in terms of its distribution function has not been reported in the literature so far and will be presented below.

\begin{figure}
  \centering
  \includegraphics[width=.85\linewidth]{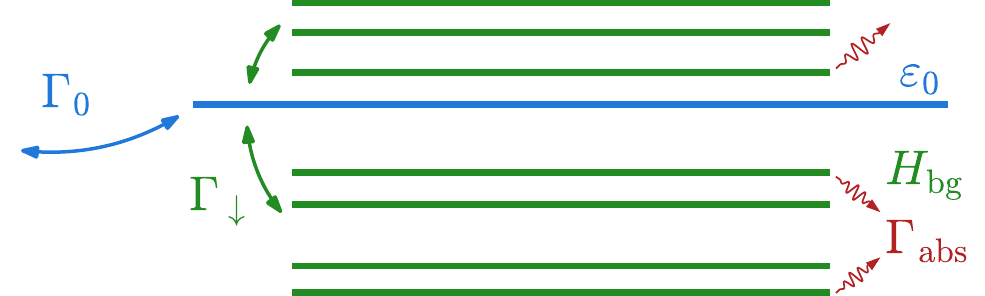}
  \caption[Schematic Hamiltonian]{
    \label{fig:schematic-Hamiltonian} %
    Schematic drawing of the model Hamiltonian.
    The single level \eps{} is coupled to scattering channels and a chaotic background
    $H_\mathrm{bg}$, leading to decay rates \Gammazero{} and  \Gammadown, respectively.
    Note that although the background states are not connected to the channels directly,
    they may have uniform broadening $\Gammaabs$ to account for homogeneous dissipation.
  }
\end{figure}

In the next section~\ref{sec:setup}, we introduce the model in detail, see Fig.~\ref{fig:schematic-Hamiltonian} for an illustration.  Section~\ref{sec:transmission} provides the derivation of the exact RMT expressions for the transmission distribution, which is the main goal of this paper, first for the ideal case of zero absorption and then in the general case of arbitrary absorption. To complete the description of scattering, we address the distribution of reflections in Sec.~\ref{sec:refl-distr}, and then summarize with the conclusion and outlook. Further details on numerical implementations as well as an extension to incorporate losses on the single level, responsible for the established transport, are given in two appendices. In all cases, a good agreement between the exact analytical results and numerical simulations with random matrices is found.

\section{Scattering setup}
\label{sec:setup}

The scattering problem in question is schematically represented by Fig.~\ref{fig:schematic-Hamiltonian}. We consider a two-port setup, which is typical for many experimental
realisations, and assume that an established transmission between two
channels occurs through a single resonance characterised by energy
\eps{} and width \Gammazero.
This resonance state, which is often referred to as a doorway state
\cite{soko97,soko92}, is coupled to a background Hamiltonian that represents
influences of a complex environment.
Due to a coupling to
surrounding complicated states, the transmission
channel spreads over the background with a rate given by the so-called
spreading width \Gammadown.
The competition between the two decay mechanism is naturally controlled
by the ratio
\begin{equation}\label{eta}
  \eta = \Gammadown/\Gammazero
\end{equation}
between the spreading and escape widths.

We restrict our consideration to systems invariant under time-reversal, which is the case most relevant experimentally. Following \cite{meh04,stoe07b}, we can then model the chaotic background by a random matrix drawn from the Gaussian orthogonal ensemble (GOE).
Additionally, we assume that the background states may have uniform broadening to account for possible homogeneous absorption in the environment. The exact RMT results for the distribution functions derived below are non-perturbative and valid at any $\eta$ and arbitrary absorption.

\subsection{Effective Hamiltonian approach}

The scattering approach based on the effective non-Hermitian
Hamiltonian~\cite{mah69,sok89,fyod11ox} is well adapted to treat both
transport and spectral characteristics on equal footing.
Neglecting global phases due to potential scattering, the scattering
matrix $S$ can be written as follows
\begin{align}
  \label{eq:S_general}
  S(E) = 1 - i A^T\frac{1}{E - \Heff} A \,,
\end{align}
where $A$ is the energy-independent matrix of the coupling amplitudes between the channel and internal states, and
\begin{align}
  \label{eq:heff_general}
  \Heff = H - \frac{i}{2} AA^T
\end{align}
defines the effective Hamiltonian of the open system. The Hermitian part $H$ corresponds to the Hamiltonian of the closed system, whereas the anti-Hermitian part accounts for finite lifetimes of resonances (eigenvalues of \Heff{}). The factorised structure of the latter ensures the unitarity of $S$ (at real scattering energy $E$). For systems invariant under time-reversal, $H$ and $A$ can be chosen as real, thus $S$ is a symmetric matrix.

We begin with quantifying a stable transmission between two channels in a ``clean'' system and consider a single level without any chaotic background. This amounts to setting above \(H = \eps\) and $A=(a\  b)$, with two real parameters \(a\) and \(b\) determining the strength of coupling to the channels. The $S$ matrix elements are then given by a multichannel Breit-Wigner formula \cite{mah69}
\begin{align}\label{S(E)}
  S(E) = 1 - \frac{i}{E-\eps+(i/2)\Gammazero} \left(
    \begin{matrix}
      a^2 & ab \\
      ab & b^2
    \end{matrix} \right),
\end{align}
where the width \Gammazero{} is given by the sum
\begin{align}\label{Gamma_0}
  \Gammazero = a^2 + b^2
\end{align}
of the partial decay widths. The peak transmission is achieved at the scattering energy $E=\eps$. The $S$ matrix evaluated at this point can be parameterised as follows
\begin{equation}\label{S_0}
   S_0  = \left(\begin{matrix}
      -r_0 & t_0 \\
      t_0 & r_0
    \end{matrix} \right),
\end{equation}
with the reflection and transmission amplitudes being
\begin{align}\label{r0,t0}
  r_0 = \frac{a^2-b^2}{\Gammazero}\,, \qquad t_0 = -\frac{2ab}{\Gammazero}\,.
\end{align}
In view of Eq.~(\ref{Gamma_0}), they satisfy the flux conservation $r_0^2+t_0^2=1$. Thus, we can express the scattering observables in terms of the experimentally measurable quantities: $\eps$, $\Gammazero$, and the transmission coefficient of the simple mode
\begin{align}
  \label{eq:def-T_0}
  T_0 \equiv t_0^2 = 1 - r_0^2\,.
\end{align}

In order to incorporate a complex environment acting on the transmission state, we follow the spreading width model \cite{Bohr,soko97} and represent the Hamiltonian as follows
\begin{align}
  \label{H_total}
  H = \left(\begin{matrix}
      \eps & \vec{V}^T \\
      \vec{V} & \HGOE
    \end{matrix}\right).
\end{align}
Here, \HGOE{} stands for the background Hamiltonian modelled by a random GOE matrix of size $N$, whereas vector $\vec{V}^T = (V_1, \dots, V_N)$ is responsible for coupling to the background. The elements of $\vec{V}$ can be chosen as real fixed or random mutually independent Gaussian variables with zero mean and a given second moment \(\langle{V^2}\rangle\). By virtue of the invariance of the GOE with respect to basis rotations, the two choices become equivalent in the RMT limit of $N\gg1$, with the obvious correspondence  $V^2=\frac{1}{N}\|\vec{V}\|^2=\langle{V^2}\rangle$. For the sake of simplicity, we will assume $\vec{V}$ fixed in the following.

Neglecting a direct coupling of the GOE states to the channels, the coupling matrix $A$ takes the special form
\begin{align}
  \label{eq:channel-matrix}
  A^T = \left(\begin{array}{cccc}
              a & 0 & \cdots & 0 \\
              b & 0 & \cdots & 0
              \end{array}\right).
\end{align}
As a result, one can easily find using Schur's complement that the $S$ matrix retains the same structure of Eq.~(\ref{S(E)}), where the following substitution is to be made
\begin{align}
  \label{res_shift}
  \frac{1}{E-\eps + (i/2)\Gammazero} \to
  \frac{1}{E-\eps + (i/2)\Gammazero - g(E)}
\end{align}
and the scalar function $g(E)$ is defined by
\begin{align}
  \label{eq:strength-function}
  g(E) &= \vec{V}^T\frac{1}{E-\HGOE}\vec{V}.
\end{align}
This is the so-called strength function \cite{Bohr}. By construction, it has the meaning of the local Green's function of the complex background \cite{fyo04}, characterising its spectral properties. When averaged over this fine energy structure, the scattering amplitudes acquire an extra damping
\begin{align}
  \label{eq:Gamma-down}
  \Gammadown{} = 2\pi V^2/\Delta
\end{align}
in addition to \Gammazero{}. The spreading width (\ref{eq:Gamma-down}) is simply Fermi's golden rule expressing the rate of decay into the ``sea'' of background states, the density of which is determined by the mean level spacing $\Delta$ \cite{Bohr,soko97}.

\subsection{$S$ matrix fluctuations}

We are interested in fluctuations in scattering at the resonance energy \eps{}. The $S$ matrix at this point can be represented in the following convenient form
\begin{align}\label{S}
  S\equiv S(\eps{}) = 1 - \frac{1}{1+i\eta K} (1-S_0),
\end{align}
where $\eta$ is given by Eq.~(\ref{eta}) and we have introduced
\begin{align}\label{K-matrix}
  K \equiv \frac{2}{\Gammadown}g(\eps{})
  = \frac{\Delta}{\pi V^2}\vec{V}^T\frac{1}{\eps-\HGOE}\vec{V}.
\end{align}
This quantity plays the role of the Wigner reaction matrix associated with scattering on the background [cf. Eq.~(\ref{S_bg}) below]. Without loss of generality, we may let \eps{} be the centre of the semicirle law determining the mean density of the background states. With the definition $\Delta^{-1}=-(1/\pi)\mathrm{Im}\left\langle\mathrm{tr}(\eps+i0-\HGOE)^{-1}\right\rangle$,
one readily finds the average value $\langle{K}\rangle=-i$, resulting in
\begin{equation}\label{<S>}
  \left\langle{S}\right\rangle = \frac{\eta}{1+\eta}+\frac{1}{1+\eta}S_0
\end{equation}
for the average $S$ matrix. This expression shows that $\eta$ controls the weight between the equilibrated and deterministic parts in scattering, which are given by the first and second terms in Eq.~(\ref{<S>}), respectively.

The average $S$ matrix is clearly non-diagonal because of channel mixing induced by $S_0$. Usually, the starting point of RMT applications to scattering ~\cite{ver85a,fyod11ox} consists in eliminating such direct processes by means of a special similarity transform \cite{enge73,nis85}. In the present case, however, this would remove the effect we are after. For similar reasons, one cannot use recent exact results \cite{kuma13,nock14} for the distribution of off-diagonal $S$ matrix elements (derived assuming a diagonal $\langle{S}\rangle$). In contrast, the obtained representation (\ref{S}) enables us to solve the problem in its full generality by applying the non-perturbative theory for the local Green's function, $K$, developed by Fyodorov, Sommers and one of the present authors in \cite{fyo04,sav05,fyod05}.

\subsection{Background as a source of dephasing}
\label{sec:dephasing}

It is instructive to give a physical interpretation to the above results in terms of the interference between the two scattering phases, the constant one due to the direct transmission and the random one induced by the chaotic background.

The deterministic part of the scattering matrix, $S_0$, can be brought to the diagonal form $O_{\varphi}^TS_0O_{\varphi}=\mathrm{diag}(-1,1)$ by an orthogonal matrix $O_{\varphi}$ that corresponds to a rotation by the angle $\varphi = \arctan\frac{t_0}{1+r_0}$. This angle expresses the degree of channel non-orthogonality due to the non-diagonal $A^TA$. By construction, the same transformation diagonalizes the full $S$ matrix (\ref{S}), yielding
\begin{align}
  \label{eq:diagonal-S-NisWei1985}
  S = O_{\varphi} \mathrm{diag}(-S_\mathrm{bg},1)O_{\varphi}^T,
\end{align}
where $S_\mathrm{bg}$ stands for the background contribution
\begin{align}\label{S_bg}
  S_\mathrm{bg} = \frac{1-i\eta K}{1+i\eta K}
\end{align}
into the full scattering process. Expression (\ref{S_bg}) is a usual form for the elastic (single-channel) scattering in open chaotic systems \cite{ver85a,sok89}, with $\eta$ now playing the role of a degree of system openness. The resulting scattering pattern is therefore due to the interference between the deterministic phase $\varphi$ and the random phase $\theta=\mathrm{arg}(S_\mathrm{bg})$, the distribution of which is well-known \cite{frie85,savi01}.

The model exemplifies the chaotic background as a natural source of dephasing in scattering processes (see also the relevant discussion in \cite{soko10}). In contrast to two other dephasing models \cite{buet86,brou97c}, our formulation is very flexible in accommodating physically relevant properties of complex environments. In particular, homogeneous losses can be easily taken into account by uniform broadening $\Gammaabs$ of the background states. Operationally, such a damping is equivalent to the purely imaginary shift $\eps{}+(i/2)\Gammaabs$ in the Green's function (\ref{K-matrix}) \cite{sav03b}. As a result, the latter becomes complex,
\begin{equation}\label{K}
  K = u - iv,
\end{equation}
with the negative imaginary part, $v>0$ (the local density of states) \cite{fyo04}. The universal statistical properties of mutually correlated random variables $u$ and $v$ are solely determined by the (dimensionless) absorption rate
\begin{equation}\label{gamma}
  \gamma = 2\pi \Gammaabs/\Delta.
\end{equation}
Their joint distribution function is known exactly \cite{sav05,fyod05} and will be applied below to study fluctuations of $S$.

\section{Transmission distribution}
\label{sec:transmission}

It is convenient to define the re-scaled transmission coefficient $T = |S_{12}|^2/T_0$, expressed in the units of the peak transmission in the ``clean'' system. We now derive the exact results for the transmission distribution function
\begin{align}
  \label{PT_def}
  \Pgamma(T) = \left\langle\delta(T - |S_{12}|^2/T_0)\right\rangle,
\end{align}
first for the ideal case of the stable background ($\gamma=0$) and then for the general case of finite absorption.

\subsection{Stable chaotic background}

\begin{figure}
  \centering
  \includegraphics[width=0.95\linewidth]{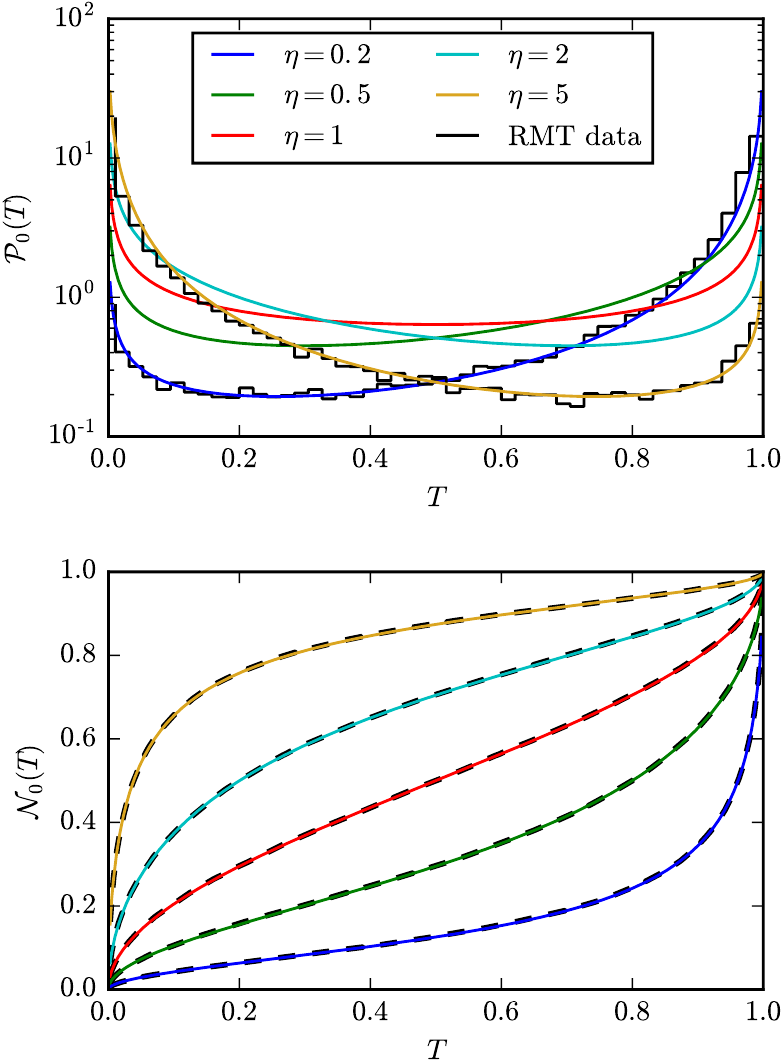}
  \caption{\label{fig:PT_without_absorbtion} %
    (color online). Transmission distribution for a stable chaotic background. The
    probability density (top) and cumulative distribution function
    (bottom) are shown at several values of the coupling parameter
    $\eta=0.2, 0.5, 1, 2, 5$. The solid curves stand for the analytical
    results (\ref{PT_stable}) and (\ref{NT_stable}). The histograms
    correspond to  numerical simulations with $5{\times}10^4$
    realizations of random \(200\times 200\) GOE matrices. The data are
    shown only for $\eta=0.2$ and $5$, the agreement for the other
    values being similar.
    The numerics for the cumulative distribution (bottom, dashed lines) is
    indistinguishable from the theory within the linewidth.}
\end{figure}

In the case of zero absorption, $K=u$ is real so the transmission coefficient is found from Eq.~(\ref{S}) as follows
\begin{align}
  \label{eq:T-of-u}
  T = \frac{1}{1+\eta^2u^2}.
\end{align}
The random variable $u$ is known \cite{mell95,fyo04} to have the standard Cauchy distribution. This stems from the fact that the scattering phase $\theta$, see Eq.~(\ref{S_bg}),
is distributed uniformly at special coupling $\eta=1$~\cite{note_phase}. The transmission distribution (\ref{PT_def}) follows then by a
straightforward integration:
\begin{align}
  \mathcal{P}_0(T)
  &= \int\limits_{-\infty}^{\infty}\frac{du}{\pi}
     \frac{1}{1+u^2} \delta \left(T - \frac{1}{1 + \eta^2u^2}\right) \nonumber
  \\
  \label{PT_stable}
  &=  \frac{1}{\pi\sqrt{T (1 - T)}} \frac{\eta}{1 + (\eta^2 - 1)T},
\end{align}
for $0\leq{T}\leq{1}$. The corresponding cumulative distribution function is given by
\begin{align}\label{NT_stable}
  \mathcal{N}_0(T)
  = 1 - \frac{2}{\pi}\arctan\left(
    \frac{1}{\eta}\sqrt{\frac{1 - T}{T}}\right).
\end{align}
Both functions are represented on Fig.~\ref{fig:PT_without_absorbtion}

The distribution (\ref{PT_stable}) has a bi-modal shape  with a square-root singularity at both edges, which is typical for transmission problems \cite{bee97}. The mean value is readily found to be  $\langle{T}\rangle = (1+\eta)^{-1}$, in agreement with the general result (\ref{<S>}), and the transmission variance reads
\begin{align}\label{var_T}
  \langle{T^2}\rangle - \langle{T}\rangle^2 = \frac{\eta}{2(1+\eta)^2}.
\end{align}
The background coupling, $\eta$, controls the weight of the distribution that is concentrated near $T\sim1$ or $T\sim0$ at small or large $\eta$, respectively. It becomes symmetric at $\eta=1$, with the variance attaining its maximum value. Further increase of $\eta$ leads to a sharp redistribution towards low transmission. For applications, this sets the limit $\eta=1$ on coupling for reliable signal transmission.

Figure~\ref{fig:PT_without_absorbtion} illustrates the above discussion and results. In order to check the validity of the predictions we present them together with numerical data from RMT simulations based on \realizations{} realizations of \(\HGOE\) blocks of
size \(200\times 200\).
The parameters \(a, b, V\), see Eqs.~(\ref{r0,t0})--(\ref{H_total}), are chosen to give \(T_0 = 0.8\) and the various \(\eta\) values. (For simplicity, we took the elements of the constant vector $\vec{V}$ to be all equal.) The overall agreement is flawless.

\subsection{Background with absorption}
\label{sec:absorption}

When the absorption rate $\gamma>0$, the complex $K$ is given by Eq.~(\ref{K}), yielding the transmission coefficient
\begin{align}\label{T}
  T = \frac{1}{(1+\eta v)^2+\eta^2u^2}.
\end{align}
The random variables $u$ and $v>0$ are mutually correlated and have the following joint distribution \cite{fyo04}:
\begin{align} \label{Puv}
  P(u,v) = \frac{1}{2\pi v^2}\Pzero\!\left(\frac{u^2+v^2+1}{2v}\right).
\end{align}
Function $\Pzero(x)$, $x=(u^2+v^2+1)/2v>1$, is known exactly at any $\gamma$ \cite{fyod05,sav05} and has the meaning of the distribution of reflection originated from the chaotic background. The parameter $x$ represents the background reflection coefficient $|S_\mathrm{bg}|^2=\frac{x-1}{x+1}<1$. Note that $S_\mathrm{bg}$ is subunitary at
finite absorption, resulting in subunitary $S$ as well.

The derivation of the transmission distribution in this case proceeds as follows. In order to perform the integration over (\ref{Puv}), it is convenient to first choose the new integration variable $y=\eta^2u^2$. With the definitions (\ref{PT_def}) and (\ref{T}), this results in
\begin{align}
  \Pgamma(T)
  &= \frac{1}{2\pi\eta T^2} \int\limits_{0}^{\infty}\frac{dv}{v^2}
     \int\limits_{0}^{\infty}\frac{dy}{\sqrt{y}}
     \Pzero\!\left(\frac{u^2 + v^2 + 1}{2v}\right)
  \nonumber \\
  &\phantom{=}\times \delta(y+(1 + \eta v)^2 - T^{-1}).
\end{align}
The $y$ integration is removed by the $\delta$ function, which restricts the remaining integration over $v$ to the domain $T^{-1}-(1+\eta v)^2 =\eta^2(v_{-}-v)(v_{+}+v)>0$, with
\begin{equation}\label{v_pm}
  v_{\pm} = \frac{1}{\eta}\frac{1\pm\sqrt{T}}{\sqrt{T}}.
\end{equation}
As a result, we arrive at the following expression
\begin{align}
  \Pgamma(T)
  = \frac{1}{2\pi\eta^2 T^2} \int\limits_{0}^{v_{-}}\frac{dv}{v^2}
    \frac{\Pzero\!\left(\frac{1+\xi^2}{2v}-\frac{1}{\eta}\right)
          }{\sqrt{(v_{-}-v)(v_{+}+v)}},
\end{align}
where the shorthand $\xi^2\equiv v_{+}v_{-}=\frac{1}{\eta^2}\frac{1-T}{T}$ has been introduced. It is now useful to choose $p =v_{-}/v-1$ as a new integration variable, yielding
\begin{align}
  \Pgamma(T) \nonumber
  &= \frac{1}{2\pi\eta^2T^2v_{-}\xi}\int\limits_{0}^{\infty}
     \frac{dp\,(1 + p)}{\sqrt{p\bigr[p + 2/(1{+}\sqrt{T})\bigl]}}
  \\
  \label{Pgam}
  &\phantom{=}\times \Pzero\!\left(\frac{1 + v_{-}^2+p(1 + \xi^2)}{2v_{-}}\right)
\end{align}
With an explicit formula for \Pzero{} found in Ref.~\cite{sav05}, representation (\ref{Pgam}) solves the problem exactly at arbitrary $\gamma$ and constitutes one of the main results of the paper.

Further analytical progress is possible in the physically interesting cases of weak and strong absorption, since the function \Pzero{} simplifies to the following limiting forms \cite{fyo04}:
\begin{align}\label{Pzero_lim}
  \Pzero(x) \approx
  \left\{
    \begin{array}{ll}
      \frac{2}{\sqrt{\pi}} \left(\frac{\gamma}{4}\right)^\frac{3}{2} \sqrt{x+1}
      \,e^{-\frac{\gamma}{4}(x+1)}, & \gamma\ll1
      \\[2ex]
      \frac{\gamma}{4}e^{-\frac{\gamma}{4}(x-1)}, & \gamma\gg1
    \end{array} \right.\,.
\end{align}
For weak absorption, $\gamma\ll1$, a close inspection of Eq.~(\ref{Pgam}) shows that the dominant contribution to the integral comes from large $p\sim1/\gamma\gg1$. In the leading order, one can neglect $p$-independent terms in the integration measure that thereby becomes ``flat''. The integration can then be performed making use of the limiting expression for \Pzero{} at small $\gamma$ stated above. This results in the following leading-order correction
\begin{equation}\label{Pgam_weak}
  \mathcal{P}_{\gamma\ll1}(T)\approx\mathcal{P}_{0}(T)
  \exp\left[-\frac{\gamma(1+(\eta-1)\sqrt{T})^2}{8\eta\sqrt{T}(1-\sqrt{T})}\right]
\end{equation}
to the zero-absorption distribution (\ref{PT_stable}), $\mathcal{P}_{0}(T)$.  Therefore, finite absorption modifies a typical bimodal shape of the transmission distribution by inducing exponential cutoffs at both edges $T\to0$ and $T\to1$.

In the opposite case of strong absorption, $\gamma\gg1$, the integral (\ref{Pgam}) is dominated by small $p\sim1/\gamma\ll1$. Performing a similar analysis as above but with the large-$\gamma$ form of \Pzero{} leads to the following approximation
\begin{align}
  \mathcal{P}_{\gamma\gg1}(T)
  &\approx
   \frac{\sqrt{\gamma\eta}(1+\sqrt{T})}{4\sqrt{\pi}(1-T)T^{3/4}\sqrt{1+(\eta^2-1)T}}
  \nonumber \\  \label{Pgam_strong}
  &\phantom{=}\times
  \exp\left[-\frac{\gamma(1-(\eta+1)\sqrt{T})^2}{8\eta\sqrt{T}(1-\sqrt{T})}\right].
\end{align}
This expression features the same exponential cutoffs at the edges, but
the bulk of the distribution gets more distorted as compared to the weak
absorption limit \cite{note_appr}.

Particularly interesting is the case of the ``critical'' coupling $\eta=1$, when the transmission distribution (\ref{PT_stable}) at zero absorption is symmetric with respect to $T\to(1-T)$. By comparing expressions (\ref{Pgam_weak}) and (\ref{Pgam_strong}), we see that such a symmetry is largely retained at weak absorption and severely violated at strong absorption, when high transmission becomes heavily suppressed.

\begin{figure}
  \centering
  \includegraphics[width=.95\linewidth]{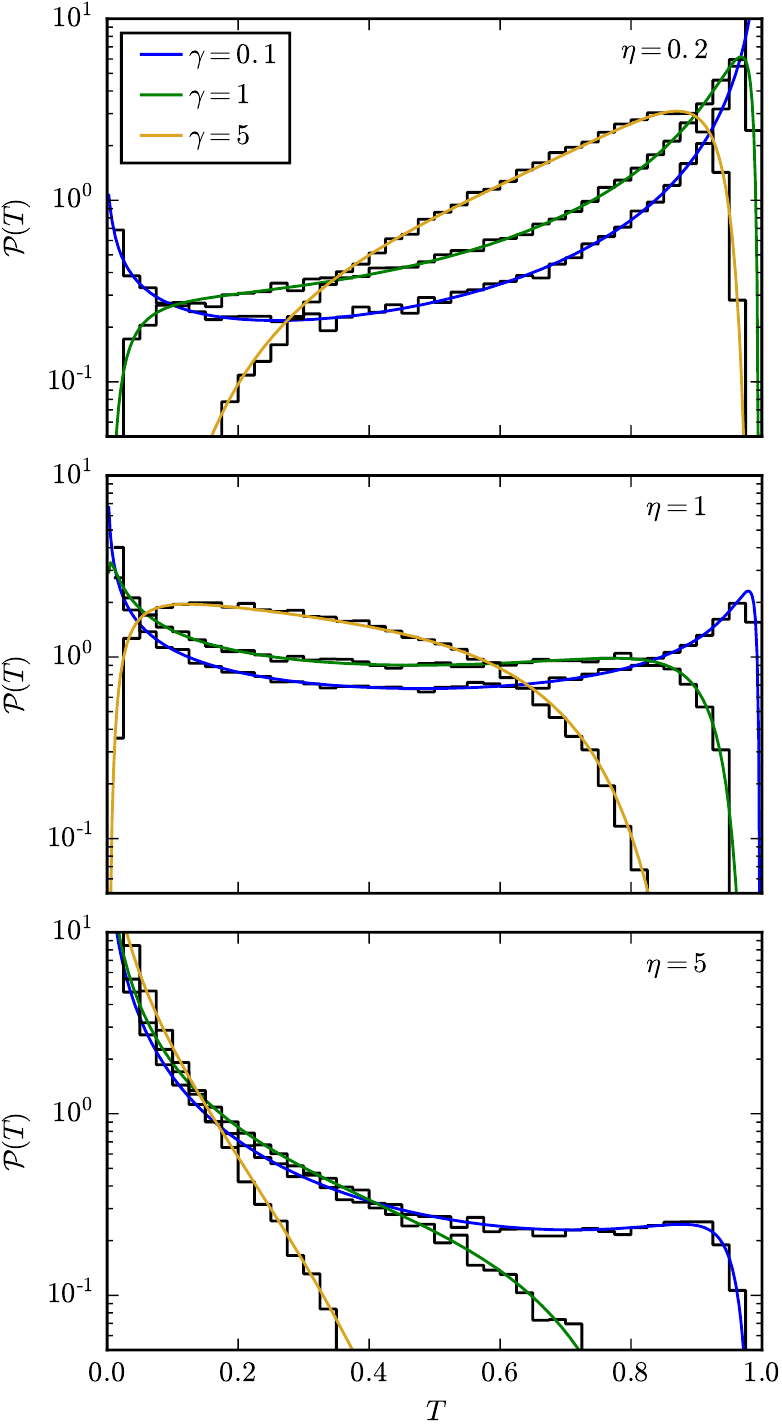}
  \caption[Transmission probability without and with
  absorption]{ \label{fig:PT_with_absorbtion} %
    (color online). Transmission distribution for a chaotic background at finite absorption.
    The curves in color correspond to Eq.~\eqref{Pgam} and are shown at
    the absorption rate $\gamma=0.1$ (blue), $\gamma=1$ (green) and $\gamma=5$ (yellow).
    The coupling to the chaotic background is chosen to be $\eta=0.2$, $1$ and $5$
    from top to bottom. Histograms stand for numerical simulations, as detailed
    in Fig.~\ref{fig:PT_without_absorbtion}. Notice the disappearance
    of the bi-modal shape of the distribution at moderate absorption.
  }
\end{figure}

At arbitrary values of $\gamma$, function \Pzero{} is given by a fairly complicated expression and the transmission distribution~\eqref{Pgam} needs to be studied numerically, see Appendix~\ref{app:interpolation} for further discussion. The corresponding results are represented on Fig.~\ref{fig:PT_with_absorbtion} for three different values of the absorption rate \(\gamma = 0.1, 1,\) and \(5\).
The aforementioned bi-modal form of distribution \(\Pgammazero\) vanishes with
increasing absorption. For the weakly coupled background, $\eta\ll1$, one observes first the diminishing of high transmission peak at $T \sim 1$, and then the eventual depletion of the peak at $T\sim0$. The situation changes at moderately strong coupling $\eta\gtrsim1$, when large transmissions become fully suppressed. Figure~\ref{fig:PT_with_absorbtion} illustrates such a behaviour, showing also the results of numerical simulations which match the analytical prediction~\eqref{Pgam} perfectly.

Note that the absorption in the numerical RMT calculations is realized by adding a constant imaginary part to the energy levels of the background. We have also checked the
agreement by modelling absorption using fictitious channels as discussed in \cite{brou97c,sav03b}.
In this case, one has to rescale the parameters \(a\) and \(b\) in order to avoid introducing losses on the established channel. The corresponding rescaling is presented in Appendix~\ref{app:rescale-params}.

\section{\label{sec:refl-distr}Reflection distributions}

The fluctuations in reflection can be studied in a similar way as for the transmission.
In the case of vanishing absorption, $\gamma = 0$, the reflection distribution can actually be related to Eq.~\eqref{PT_stable}. Due to the unitarity of $S$ in this case, we have
$|S_{11}|^2 = 1-|S_{12}|^2=|S_{22}|^2$. Therefore, the distribution of the reflection coefficient
$R_{c}=|S_{cc}|^2$ ($c=1,2$) is determined by Eq.~\eqref{PT_stable} according to
\begin{align}
  \label{eq:Prefl1-no-absorb}
  \Pgammazero^{\mathrm{(refl)}}(R_{c}) = \frac{1}{T_0}\Pgammazero
  \left(\frac{1 - R_{c}}{T_0}\right)
\end{align}
in the region $1-T_0\leq R_c\leq1$, being zero otherwise.

In the case of finite absorption, transmission and reflection coefficients are no longer related by flux conservation. The reflection coefficients are readily found from Eqs.~\eqref{S} and \eqref{K} in the explicit form
\begin{equation}
  \label{R_1,2}
  R_{1,2} = \frac{(\eta v\mp r_0)^2+\eta^2u^2}{(\eta v+1)^2+\eta^2u^2},
\end{equation}
where the upper (lower) sign stands for $R_1$ ($R_2$).
One sees that the reflection coefficients are generally different at nonzero $r_0$.
This is a manifestation of the interference between the equilibrated and direct reflection induced by $S_0$ (see the discussion in Sec.~\ref{sec:dephasing}).

Following similar steps as leading to Eq.~\eqref{Pgam}, we arrive at the
following representation for the distribution of the reflection
coefficient $R_1$ at \(\gamma > 0\):
\begin{align}
  \Prefletagamma(R_{1})
  &=
    \frac{(1+r_0)}{2\pi\eta^2(1 - R_{1})^2}
    \int\limits_{0}^{\infty}
    \frac{dp\,(1 - r_0 + 2 \eta w(p))}{\sqrt{p}\,(\hat{w}_+ + p w_-)^2}
    \notag
    \\
  \label{eq:pR_absorb}
  &\phantom{=}\times
    \frac{(w_+ + w_-)}{\sqrt{y(p)}}
    \Pzero\left(\frac{1 + w(p)^2 + py(p)}{2w(p)}\right).
\end{align}
Here, we have introduced the following shorthand notations:
$\hat{w}_\pm = \frac{1}{\eta}\frac{r_0\pm\sqrt{R_1}}{1 \mp\sqrt{R_1}}$,
$w_\pm = \max\{0, \hat{w}_\pm\}$,
$w(p) = {(\hat{w}_+ + p w_-)}/{(p + 1)}$ and
$y(p) = (\hat{w}_+ - w_-)[(\hat{w}_+ - \hat{w}_-) - p(\hat{w}_- - w_-)]/{(p + 1)^2}$. The distribution of $R_2$ is given by the same
formula (\ref{eq:pR_absorb}) under the replacement $r_0\to-r_0$
everywhere there.

\begin{figure*}
  \centering
  \includegraphics[width=0.95\linewidth]{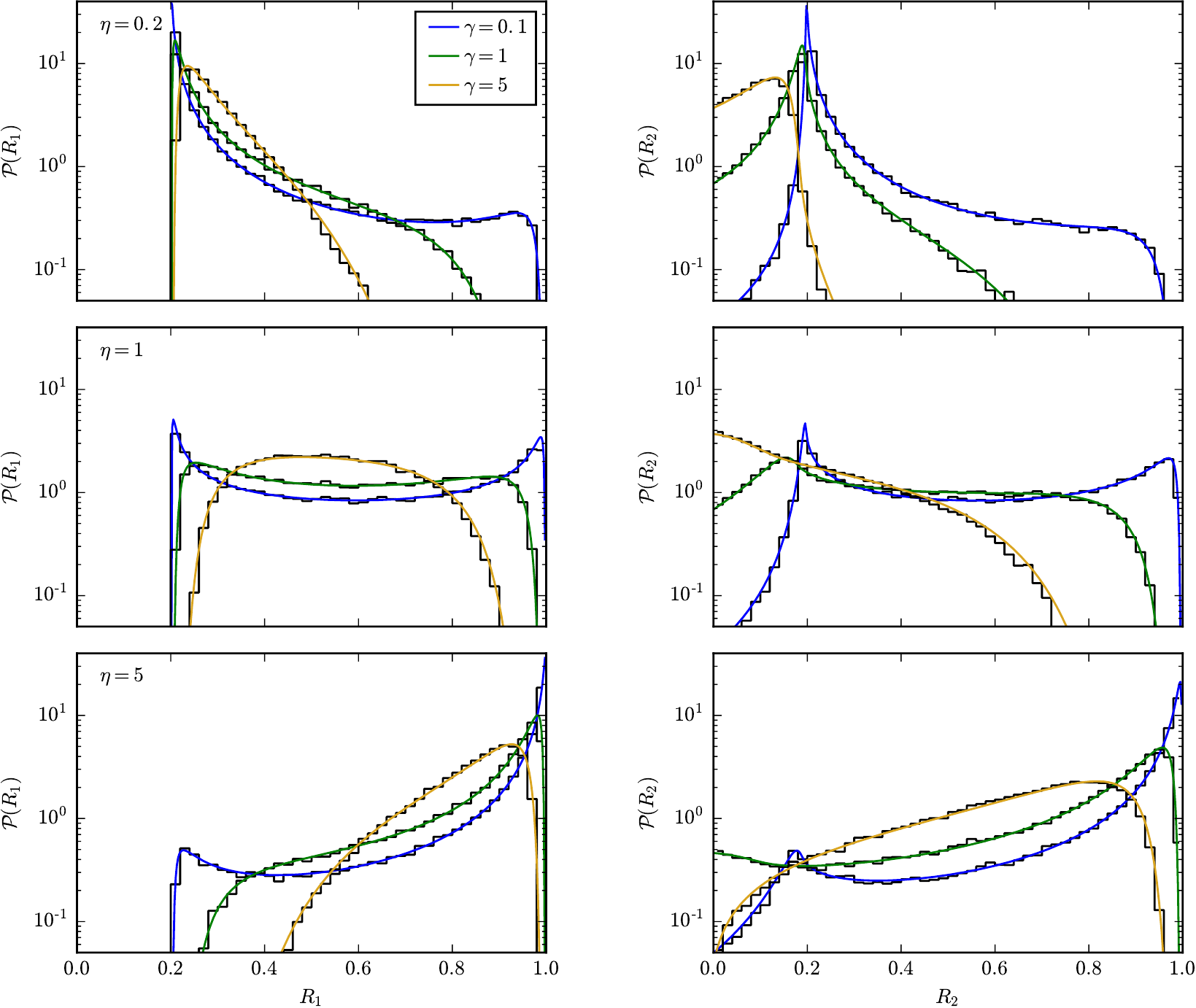}
  \caption[Reflection at finite absorption]{ \label{fig:PR_with_absorbtion} %
    (color online).
    Distribution of reflection coefficients $R_1$ (left panel) and $R_2$ (right panel),   where $R_{c}=|S_{cc}|^2$. The curves in color correspond to Eq.~\eqref{eq:pR_absorb}
    and are shown at the different absorption rates $\gamma=0.1,1$, and $5$
    (in blue, green and yellow, respectively) and background coupling $\eta=0.2$, $1$ and $5$
    (from top to bottom). The numerical results of RMT simulations are shown as
    histograms (the same statistics was used as for Fig.~\ref{fig:PT_with_absorbtion}).
    The chosen deterministic transmission \(T_0=0.8\).
}
\end{figure*}

Due to the $r_0$-dependence of expression~\eqref{eq:pR_absorb}, it follows that there is a particular difference in the distributions of the two reflection coefficients. Without loss of generality, one can choose $r_0<0$ corresponding to the unequal channel couplings, $|b|>|a|$, see Eq.~(\ref{r0,t0}). Then we have
\begin{align}
  \label{eq:wplus-R11}
  w_+ + w_- = \left\{\begin{matrix}
      0, & & R_1 < r_0^2 \\
      \hat{w}_+, & & R_1 \ge r_0^2
    \end{matrix}\right.
\end{align}
for the reflection coefficient \(R_1\) and
\begin{align}
  \label{eq:wplus-R22}
  w_+ + w_- = \left\{\begin{matrix}
      \hat{w}_+, & &R_2 < r_0^2 \\
      \hat{w}_+ + \hat{w}_-, & & R_2 \ge r_0^2
    \end{matrix}\right.
\end{align}
for \(R_2\).
This implies that \(\Prefletagamma(R_1) = 0\) identically at $R_1 \le r_0^2 = 1-T_0$, leading to the same gap for small reflections as seen from the case without absorption. The distribution of the other reflection coefficient \(R_2\) (i.e., in the channel with stronger coupling) does not have such a gap since expression (\ref{eq:wplus-R22}) is nonzero for all $R_2$.

In Fig.~\ref{fig:PR_with_absorbtion} we present a comparison of the analytical result \eqref{eq:pR_absorb} with numerical data, using the same choice of the parameters as for the transmission distribution before. In the general case of finite backscattering, $r_0\neq 0$, the reflection distributions in two channels show a distinctly different behavior as discussed above. We have chosen the value of $r_0=-\sqrt{0.2}$, corresponding to the established transmission \(T_0 = 0.8\). Therefore, the behavior of the reflection coefficient \(R_{2}\) changes
below \(1 - T_0 = 0.2\) where the gap in the distribution vanishes. The overall agreement of the theory with numerics is flawless.

\section{Conclusions and Outlook}

In this work, we have formulated an approach to characterise fluctuations in an established transmission that are induced by a chaotic background. Our method is based on the strength function formalism, adopted from and developed in nuclear physics, providing new insights for the applications of the latter in a broader context of wave chaotic systems. The strength of coupling to the background is controlled by the single parameter (\ref{eta}), the ratio of the spreading to escape width. Using RMT to model the chaotic background, we have derived the transmission distribution in an exact form valid at arbitrary uniform absorption in the background. The analytical results are supported by extensive numerics performed by Monte-Carlo simulations with random matrices.

The distribution has a bimodal shape, with two peaks at low and high transmission that are exponentially suppressed at finite absorption. It takes simple limiting forms in the physically interesting cases of weak and strong absorption, which we have discussed in detail as well. Fluctuations in high transmission are found to be affected more strongly by finite absorption, when the background coupling exceeds certain limiting value. These results may be relevant in the reliability context of wireless communication devices \cite{cou11}.

The method developed is very flexible in incorporating physical properties of the system. In particular, we have neglected absorption of the transmission line itself, however, the latter can be important for experimental realisations, e.g., including ongoing research with chaotic reverberation chambers. Such an extra damping can be naturally accommodated into the theory by a simple rescaling procedure as outlined in Appendix~\ref{app:rescale-params}. Following \cite{rozh04,savi06b}, one can also include effects due to nonuniform absorption in the environment. The method can be generalised to other types of the chaotic background (e.g., without time-reversal) as well as to multichannel transmission, where the complete characterisation of both reflection and total transmission in terms of their joint distribution is actually possible and will be reported elsewhere~\cite{savi17}. Therefore, we expect our results to find further applications in studying wave propagation with complex environments.

\section*{Acknowledgments}

Three of us (D.V.S., M.R.\ and U.K.) would like to acknowledge a stimulating environment during the XII Brunel-Bielefeld Workshop on Random Matrix Theory and Applications held on 9--10 December 2016 at Brunel, UK, where the work along the lines presented above was initiated. Partial financial support by Horizon 2020 the EU Research and Innovation Program under grant no. 664828
(NEMF21~\cite{nemf21}) is acknowledged with thanks.

\appendix
\section{\label{app:interpolation} Accuracy of the interpolation formula}

In order to ease both implementation and analytical treatment of the exact expressions, one can
use a much simpler interpolating formula for function \Pzero{}, which was suggested in~\cite{fyo04}:
\begin{equation}\label{Pzero_int}
  \Pzeroint(x) = C_\gamma^{-1}
  \bigl(A_{\gamma}\sqrt{\gamma(x+1)} + B_\gamma\bigr) e^{-\frac{\gamma}{4}(x+1)}.
\end{equation}
Here, $A_\gamma=(e^{\gamma/2}-1)/2$,
$B_\gamma=1+\frac{\gamma}{2}-e^{\gamma/2}$, and the normalisation
constant $C_\gamma = \frac{4}{\gamma} (2\Gamma(\frac{3}{2},\frac{\gamma}{2})A_\gamma+e^{-\gamma/2}B_\gamma)$,
with $\Gamma(\nu,\alpha)$ being the upper incomplete gamma function.
This formula was earlier found to work surprisingly well when compared to the exact result~\cite{sav05}. Here we show that the level of agreement with the full
implementation of the exact \Pzero{} yields equally good results.

\begin{figure}[b]
  \centering
  \includegraphics[width=.95\linewidth]{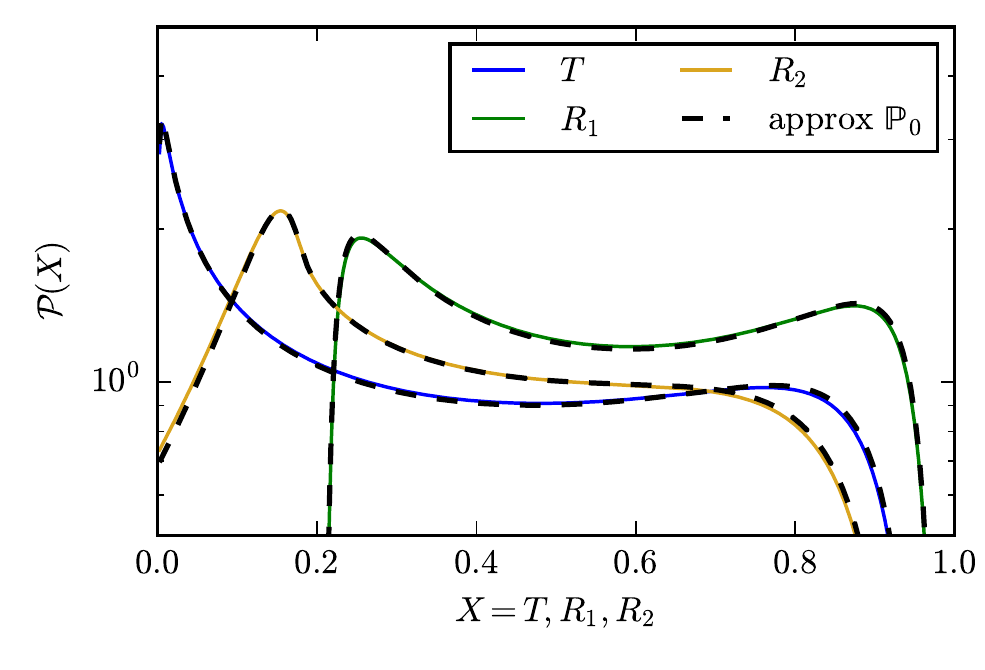}
  \caption[Comparison for approximate \Pzero]{
    \label{fig:P_0_approx-compare} %
    (color online). Accuracy of the interpolation formula.
    The exact distributions of $T$, $R_{1}$, and $R_{2}$ are
    shown in solid blue, green, and yellow lines, respectively.
    Shown in dashed black lines are the same functions from Eqs.~\eqref{Pgam}
    and~\eqref{eq:pR_absorb} but using the interpolation formula \Pzeroint{}
    from Eq.~\eqref{Pzero_int}. %
    All curves are for \(\eta = 1\), \(\gamma = 1\), \(T_0 = 0.8\).
    For other parameter values the correspondence is equally good.
  }
\end{figure}

Figure~\ref{fig:P_0_approx-compare} shows two types of the three analytical curves derived from Eqs.~\eqref{Pgam} and~\eqref{eq:pR_absorb}: one time calculated using \Pzero{} and one time
calculated by \Pzeroint. The biggest deviations can be found towards
\(0\) and \(1\) or around \(r_0^2\) in the reflection distributions. However, the overall accuracy of the interpolation formula is very good.

\section{\label{app:rescale-params} Absorption of the transmission line}

As mentioned in the main text, the derived distributions of transmission and reflection do not include a possible absorption of the single level \eps. The latter can be easily incorporated by shifting $\eps\to\eps-(i/2)\Gammaabs^{(0)}$ in Eqs.~(\ref{H_total}) and (\ref{res_shift}), where the absorption width $\Gammaabs^{(0)}$ of the transmission line is generally different from that of the background. This amounts to replacing $\Gammazero \to \Gammazero+\Gammaabs^{(0)}$ and thereby the channel coupling constants $a$ and $b$ by
\begin{align}
  a' = a\,\sqrt{1 + \Gammaabs^{(0)}/\Gammazero}, \quad
  \label{eq:rescale-a-b}
  b' = b\,\sqrt{1 + \Gammaabs^{(0)}/\Gammazero}.
\end{align}
This results in the following rescaling of the parameters
\begin{align}
  T_0 &\to T'_0 = T_0 \left(1+\Gammaabs^{(0)}/\Gammazero\right)^{-2}
  \\
  \eta &\to \eta'= \eta\left(1 + \Gammaabs^{(0)}/\Gammazero\right)^{-1}
\end{align}
in the expressions~\eqref{Pgam} and~\eqref{eq:pR_absorb}. Using these
variables allows us to predict the corresponding distributions also in the case of finite absorption of the transmission line.


\begin{thebibliography}{53}
\bibitem [{not({\natexlab{a}})}]{note_phase}%
  \BibitemOpen
  \bibinfo {note} {Note that the average
  $\langle{S_\mathrm{bg}}\rangle=(1-\eta)/(1+\eta)$. At $\eta=1$,
  $\langle{S_\mathrm{bg}}\rangle=0$ implies the uniform distribution for
  $\theta$ and thus the Cauchy distribution for
  $u=-\arctan\frac{\theta}{2}$.}\BibitemShut {Stop}%
\bibitem [{not({\natexlab{b}})}]{note_appr}%
  \BibitemOpen
  \bibinfo {note} {We note that although both
  expressions (\ref{Pgam_weak}) and (\ref{Pgam_strong}) are not  normalised,
  they reproduce the exact asymptotic behaviours near the edges. The accuracy
  of the approximations can be systematically improved by keeping further
  order terms in $1/p$ (or in $p$) for weak (or strong) absorption when
  performing the integration in Eq.~(\ref{Pgam}).}\BibitemShut {Stop}%
\end{thebibliography}
\end{document}